\documentclass[12pt,pra,showpacs,aps]{revtex4}
\usepackage{graphicx}
\usepackage{dsfont,psfrag}
\usepackage{amsmath,amsfonts,amssymb,latexsym,amsthm,amscd}
\usepackage{times,amsthm}
\usepackage{epsfig}
\usepackage{textcomp}
\usepackage{rotating}
\usepackage{here}
\usepackage{bm}
\usepackage[latin1]{inputenc}

\newcommand {\BS} {\begin{eqnarray}}
\newcommand {\ES} {\end{eqnarray}}

\newcommand{\vecx}{{\bm x}}
\newcommand{\veck}{{\bm k}}

\renewcommand {\eqref}[1]{(\ref{#1})}
\renewcommand {\tfrac}[2]{\text{$\frac{#1}{#2}$}}
\renewcommand {\text}[1]{\mbox{#1}}

\newcommand{\ie}{i.\thinspace{}e. }

\begin{document}

\title{The granularity of weakly occupied bosonic fields beyond the local
  density approximation}

\author{M. Eckart, R. Walser and W.P. Schleich} 
\affiliation{Institut f\"ur Quantenphysik, Universit\"at Ulm, 89069 Ulm, Germany} 
\email{michael.eckart@uni-ulm.de}
\author{S. Z\"ollner and P. Schmelcher} 
\affiliation{Theoretische Chemie, Universit\"at Heidelberg, INF 229, 69120 Heidelberg, Germany} \date{\today}

\date{\today}

\begin{abstract}
  We examine ground state correlations for repulsive, quasi one-dimensional
  bosons in a harmonic trap. In particular, we focus on the few particle limit
  $N=2,3,4,\ldots$, where exact numerical solutions of the many particle
  Schr\"odinger equation are available employing the Multi-Configuration
  Time-dependent Hartree method.  Our numerical results for the inhomogeneous
  system are modeled with the analytical solution of the homogeneous problem
  using the Bethe ansatz and the local density approximation.  Tuning the
  interaction strength from the weakly correlated Gross-Pitaevskii- to the
  strongly correlated Tonks-Girardeau regime reveals finite particle number
  effects in the second order correlation function beyond the local density
  approximation.
\end{abstract}

\pacs{
03.65.Ge, 
03.75.-b, 
05.30.Jp, 
67.85.-d 
}


\maketitle

\pagebreak

\tableofcontents

\section{Introduction}

Observing strongly correlated atomic quantum gases in situ and real time is
quite an achievement
\cite{Goerlitz01,Esslinger03,Tolra04,Paredes2004,Ertmer03,Weiss05,
  schmiedmayer,vandruten08}. By controlling the trapping geometry one can
effectively adjust the ``degree'' of dimensionality, by feeding in more
particles one can approach the thermodynamic limit, a central concept of our
macroscopic world, and by controlling the coupling constant one can switch
between universality classes of physical systems.  Maybe, all of this was once
envisioned by the great minds who have conceived the very few exactly solvable
models of many-body physics
\cite{Bethe31,Gir60,Lieb63I,richardson68,MathisLiebBuch,exactsolutions}, but
witnessing the merger of expectation and experiment proves to be an exciting
period, today.
 
In the current article we have explored a particular aspect of this rich
topic, by focusing on the quantum properties of the ground state containing
only a few bosonic particles, \ie $N=2,3,4$ inside a harmonic, one-dimensional
trap. This situation is akin to the atomic or nuclear physics limit of a
condensed matter system. There the implied granularity of fermionic matter
appears as a shell structure in the energy configuration or on energy surfaces
through the appearance of magic quantum numbers. This was unheard of in the
field of uncharged gaseous matter until recently with the experimental use of
neutral, repulsive, bosonic atoms \cite{bloch08a}.
  
A striking example here is given by the one-dimensional (1D) Bose gas. Its
closed-form solution has been given for the homogeneous system using Bethe's
ansatz in the pioneering work of Lieb \cite{Lieb63I}, which focused on the
thermodynamic limit ($N\to\infty$ at fixed density).  These results have been
extended to include finite particle numbers \cite{Sakmann05} and effects due
to a slowly varying trapping potential \cite{Petrov00,dunjko01a}, in the
sense that the thermodynamic-limit results for the homogeneous system still
hold locally even in the presence of an inhomogeneity. However, for small atom
numbers $N$, the trapped system can be solved in a numerically exact fashion,
without resort to such a local-density approximation
\cite{hao06a,Pfannkuche07,zollner2006,zollner2006a,Schmidt07}.  This is
the starting point of our paper, which aims at a detailed comparison of the
exact correlation functions of $N$ trapped bosons with those obtained in a
finite-size homogeneous system under a local-density approximation. That way
we map out intrinsic confinement effects and discuss the validity of the
local-density approximation for small atom numbers.

Following this motivation, we will present the basic model of $N$
particles trapped in one dimension in section~\ref{model}. Next, we
introduce our physical observables and computational methods in section~\ref{obs}.
The homogeneous limit of this system is the Lieb-Liniger model, which
represents our benchmark. Its basic notions will be reviewed briefly
in section~\ref{hom} and used in section~\ref{inhom} to discuss our
numerical results and analytical modeling of a few harmonically trapped
particles.


\section{Model}\label{model}
Let us consider a gas of $N$ one-dimensional bosons with repulsive, short range
interactions in a harmonic trap \cite{Pfannkuche07, Schmidt07, Giorgini03, Olshanii03}. Then, the dynamics is given by the dimensionless Hamiltonian
\begin{eqnarray} 
\label{Hamiltonian0}
H&=&  \sum\limits_{j=1}^N \left(-\tfrac{1}{2}\partial^2_{j}
+ \tfrac{1}{2} x_j^2\right)+ \sum\limits_{j<l=1}^N g\delta(x_j - x_l) \; ,
\end{eqnarray}
where we have measured energy in units of $\hbar \omega$, length in multiples
of the harmonic oscillator length $a_0=\sqrt{\hbar/m\omega}$ and used the
short hand notation $\partial_j=\partial/\partial {x_j}$. For example, such an
effective one-dimensional description can be obtained by starting with real
three-dimensional bosonic atoms of mass $m$ in a strongly anisotropic external
trapping potential \mbox{$V(\textbf{r})= \frac{1}{2}m\omega^2 x^2 +
\frac{1}{2}m\omega_\perp^2(y^2+z^2)$}.  If the transverse level spacing is much
larger than in the axial direction $\beta=\omega_\perp/\omega\gg 1$, we can
integrate out two dimensions by assuming that the two-dimensional
$(yz)$-subsystem only occupies the ground state. This procedure leads to the
quasi one-dimensional coupling constant $g=2 \beta a_s/a_0$, where $a_s$
denotes the s-wave scattering length of the bosons.

Nowadays, this situation can be realized experimentally \cite{bloch08a} and it
is possible to investigate quantum correlations in situ. In particular, we are
interested in the properties of the ground state of a few interacting bosons,
i.\,e. $N=2,3,4,\ldots,$ and we will explore their quantum correlations.  In a
homogeneous system of length $L=a_0 \ell$ with a dimensionless scale $\ell$,
the linear number density $n=N/\ell$ is translation invariant and the
qualitative behavior of the ground state strongly depends on the correlation
parameter $\gamma=g/n$. This was first described by Lieb and Liniger
\cite{Lieb63I}. In the thermodynamic limit $\lim_{N,\ell\rightarrow
  \infty}N/\ell=n$, it turns out that the state of the homogeneous gas of
bosons is completely characterized by this parameter.  It is customary to call
bosons weakly correlated for $\gamma\ll 1$ (Gross-Pitaevskii regime) and
strongly correlated for $\gamma \gg 1$ (Tonks-Girardeau regime).

The local density approximation extends this description to weakly
inhomogeneous systems under the assumption that the variation of the ground
state follows parametrically the spatial variation of the single particle
density.  In the following, this assertion will be probed by explicitly
constructing the $N$-body wave function with the Bethe ansatz. We will analyze
the behavior of the state over the whole range of interaction strengths for an
increasing particle number ($N=2,3,4$) in order to investigate the transition
towards the thermodynamic limit. The correlation functions of the ground state
strongly depend on the particle number and this effect is most significant for
few bosons. This analysis will provide a profound understanding of the
inhomogeneous system, which can only be solved numerically otherwise.

\section{Observables and Computational methods}\label{obs}
Assuming we have complete knowledge of the symmetrized and normalized
$N$-particle wave function $\Psi(\vecx)=\Psi(x_1, \ldots, x_N)$, then we need
to extract relevant information about its behavior in terms of experimentally
accessible observables \cite{peletminskii,Martin2004}.  Most relevant for this
purpose are the number density $n(x) = N \rho(x)$, which is proportional to
the single particle density \BS \rho(x) &=& \int dx_2 \cdots dx_N
|\Psi(x,x_2,\ldots,x_N)|^2 \; , \ES the first order correlation function
measuring phase coherence
\begin{eqnarray} 
\label{g1}
g^{(1)}(x,y)&=& \frac{\int dx_2 \cdots dx_N 
\Psi^*(x,x_2,\ldots,x_N)\Psi(y,x_2,\ldots,x_N)}{\sqrt{\rho(x) \rho(y)}}
\end{eqnarray}
and the diagonal of the second order correlation function measuring density
fluctuations
\begin{eqnarray} 
\label{g2}
g^{(2)}(x,y)&=& 
\frac{N-1}{N}\frac{\int dx_3 \cdots dx_N |\Psi(x,y,x_3,\ldots,x_N)|^2}{\rho(x) \rho(y)} \; .
\end{eqnarray}
We will focus on the behavior of the second order correlation function as it
is a more sensitive probe for quantum statistical correlations in the system.
Theoretically, much attention has already been directed towards second order
correlation functions \cite{Petrov00,Olshanii03, ShlyapGang03, Shlyap03,
  walser04, Bog04,Shlyap05, Giorgini05}, but the discussions were mostly
concerned with their properties in the thermodynamic limit.  In contrast, our
interest is directed towards the detailed behavior of the second order
correlation function for few boson systems, which differs from the
thermodynamic limit.

The next subsection is devoted to a brief introduction of the
Multi-Configuration Time-Dependent Hartree (MCTDH) method which is used to
calculate the $N$-body ground state of the Hamiltonian in the presence of a trap
in \eqref{Hamiltonian0}.

\subsection{Multi-Configuration Time-Dependent Hartree method}

The numerically exact MCTDH method \cite{Meyer90,Meyer00} is a
quantum-dynamics tool which has been applied successfully to systems of few
identical bosons
\cite{zollner2006,zollner2006a,zollner2007,zollner2008,zollner2008a,zollner2008b}.
It solves the time-dependent $N$-body Schr\"odinger equation $(i \hbar
\partial_t-H) {\Psi}(\vecx,t) =0$, as an initial-value problem by expanding the
solution \BS
\label{HD_ansatz}
\Psi(\vecx,t) & = & \sum\limits_{J\in \mathcal{C}} a_J(t) \Phi_J(\vecx,t), \ES
in terms of direct (or Hartree) product states $\Phi_J(\vecx,t)
=\prod_{i=1}^N\phi_{j_i}(x_i,t)$ and summing over all admissible
configurations $\mathcal{C}=\{J=(j_1,\ldots,j_N)|1\leq j_i\leq s\}$.  In turn,
the still unknown, best single-particle functions $\{ \phi_{j}(x,t)|1\leq j
\leq s\}$ are represented in a fixed primitive basis implemented on a grid and
$s$ denotes the maximum number of required basis functions.  The permutation
symmetry of $\Psi(\vecx,t)$ is ensured by the correct symmetrization of the
expansion coefficients $a_J(t)$.

Using the Dirac-Frenkel
variational principle, one can derive equations of motion for both $a_J(t)$,
and $\phi_j(x,t)$ \cite{Meyer00}. Integrating this system of
differential-equations allows us to obtain the time evolution of the system
via \eqref{HD_ansatz}. This has the advantage that the basis
$\{\Phi_J(t)|J\in\mathcal{C}\}$ is variationally optimal at each time $t$.
Thus it can be kept relatively small, rendering the procedure very efficient.

Although designed for time-dependent simulations, it is also possible to apply
this approach to stationary states. This is done via the so-called relaxation
method \cite{Kosloff86}. The key idea is to propagate an initial wave function
$\Psi(\vecx,t=0)$ by the non-unitary, imaginary time propagator
$U(\tau)=e^{-H\tau}$.  As $\tau \to \infty$, any excited state contribution is
exponentially suppressed with $e^{-(E_m - E_0)\tau}$ and we are left with the
ground state. In practice, one relies on a more sophisticated scheme termed
improved relaxation \cite{Meyer03}, which is much more robust especially for
excitations. Here, the energy $E=\langle \Psi |H|\Psi\rangle$ is minimized with
respect to both the coefficients $a_J$ and the orbitals $\phi_j(x)$.  The
effective eigenvalue problems thus obtained are then solved iteratively by
first solving for $a_J$ with fixed orbitals and then optimizing $\phi_j(x)$
by propagating them in imaginary time over a short period. That cycle will
then be repeated.

Applying this procedure, we obtain the exact wave function for the ground
state and can subsequently calculate the correlation functions according to
\eqref{g1} and \eqref{g2}. A profound understanding of the qualitative as
well as quantitative details of this correlation function is the main topic of
this contribution. For this purpose we compare the results of the MCTDH method
to Lieb-Liniger theory for the homogeneous Bose gas which will be combined
with a local density approximation. However, before doing so we briefly review
the main concepts of the Bethe ansatz and Lieb-Liniger theory, which can be
solved exactly.

\subsection{Lieb-Liniger theory and Bethe ansatz for the homogeneous system}
As we have discussed, it is only possible to obtain the eigenstates and
eigenvalues of the trapped Hamiltonian in \eqref{Hamiltonian0} with a
significant computational effort.  However if one can disregard the external
trapping potential and supply the system with periodic boundary conditions
instead, we obtain the Hamiltonian
\begin{eqnarray} 
\label{Hamiltonian_LL}
H_{LL}&=&  \sum\limits_{j=1}^N -\tfrac{1}{2} \partial^2_{j}+ 
\sum\limits_{j<l=1}^N g\delta(x_j - x_l)\; .
\end{eqnarray}

The corresponding eigenvalue problem has been solved analytically by Lieb and
Liniger \cite{Lieb63I,Sakmann05} using the Bethe ansatz \cite{Bethe31} and they
derived the ground state properties -- even for the thermodynamic limit.
Before we discuss correlation functions and their functional behavior, we
will briefly recall the quintessential steps.

To solve for the ground state and its energy, we have to consider the
eigenvalue equation \BS
\label{LL-Sch}
H_{LL} \Psi(\vecx) &=& {E} \Psi(\vecx),\ES where the spatial region is
$\{\vecx=(x_1,\ldots,x_N)| \, 0 \leq x_j < \ell\}$ and the totally symmetric
wave function satisfies periodic boundary conditions.  The dimensionless
system length $\ell$ determines the linear number density $n=N/\ell$ for a
given particle number $N$. It still occurs in our dimensionless formulation of
the Lieb-Liniger Hamiltonian because we want to allow for a straightforward
comparison of the ground states of the Hamiltonians in 
\eqref{Hamiltonian0} and \eqref{LL-Sch} for equal particle numbers $N$ and
equal interaction strengths $g$. Furthermore, having the length $\ell$ as a
free parameter we can tune the number density $n$ such that the correlation
parameter in the homogeneous system is equivalent to the correlation parameter
at a certain position in the trapped system.

A subspace of the whole configuration space is the spatial simplex that
contains only ascending coordinate $N$-tuples, i.\,e.
$\mathcal{R}=\{\vecx=(x_1,\ldots,x_N)| \,0 \leq x_1 < x_2 < \ldots < x_N <
\ell\}$.  In this region, \eqref{LL-Sch} together with its periodic
boundary conditions are equivalent to \BS
\label{LL1}
(\sum_{j=1}^{N}-\tfrac{1}{2} \partial^2_j) \Psi(\vecx) & = & E \Psi(\vecx),\\
(\partial_{j+1} -\partial_j)\Psi(\vecx)\Big|_{x_{j+1}=x_j} &=& 
g \Psi(\vecx)\Big|_{x_{j+1}=x_j},\\
\label{LL3}
\Psi(0,x_2, \ldots, x_N) & = & \Psi(x_2, \ldots, x_N,\ell), \\
\label{LL4}
\partial_x \Psi(x,x_2, \ldots, x_N)\Big|_{x=0} &=& \partial_x \Psi(x_2,
\ldots, x_N,x)\Big|_{x=\ell} \; .  \ES Due to the required symmetry of the
wave function under particle exchange, knowledge of $\Psi(\vecx)$ in the
region $\mathcal{R}$ is equivalent to knowing $\Psi(\vecx)$ in all other
regions of the configuration space. In order to solve
(\ref{LL1}-\ref{LL4}), one uses the ansatz \BS
\label{ansatz}
\Psi(\vecx) &=&\sum\limits_{\mathcal{P} \in \mathcal{S}_N} a_\mathcal{P}
\,e^{i \veck_{\mathcal{P}}\vecx} \; , \ES where the summation extends over all
$N!$ elements $\mathcal{P}$ of the permutation group $\mathcal{S}_N$.  If we
denote the wave vector of the $N$ particles by $\veck = (k_1, \ldots, k_N$),
then the permuted vector is $\veck_{\mathcal{P}}= (k_{\mathcal{P}(1)},
\ldots,k_{\mathcal{P}(N)})$. For convenience we also introduce the scalar
product $\veck_{\mathcal{P}}\vecx = \sum_{j=1}^{N}k_{\mathcal{P}(j)} x_j$.
Immediately, one obtains for the ground state energy of the $N$-particle system
\BS E &=& \tfrac{1}{2} N n^2\, e_{B}(N,\gamma), \quad e_{B}(N,\gamma) =
\frac{1}{N^3}\sum\limits_{j=1}^N (k_j \ell)^2 \; .  
\ES

The remaining task consists of determining the wave vector $\veck$ such
that all boundary conditions are fulfilled. After minor algebra which is
outlined in \cite{Lieb63I}, one obtains the following equations for the
components of the wave vectors of the ground state 
\BS
\label{quotient1}
(k_{j+1}-k_j)\ell &=& \sum\limits_{i=1}^N(\theta_{j+1 i}-\theta_{ji}) + 2\pi \quad \text{for} \quad j=1, \ldots, N-1 \; ,\\
\label{quotient2}
\theta_{rs} & =& 2 \arctan{\left[\frac{(k_s-k_r)\ell}{N \gamma}\right]} \; .
\ES  Furthermore
we note that the ground state solution possesses reflection symmetry, which means
that for every positive component $k_j$ there exists a negative counterpart
$-k_j$.

Before the ground state wave function can be calculated, we need to clarify
how the factors $a_\mathcal{P}$ are defined, such that the boundary conditions
are fulfilled. This is done by first setting $a_\mathds{1} = 1$.  If
$\mathcal{P}$ takes $\veck$ into $\veck_{\mathcal{P}}$, then this is achieved
by subsequent transpositions. For each transposition, the amplitude acquires a
factor $-e^{i\theta_{lj}}$, if $k_l$ and $k_j$ are transposed and $k_l$ is to
the left of $k_j$. The product of all these factors is $a_\mathcal{P}$.  Thus,
we end up with the following wave functions for $N=2,3,4$ particles
\BS
  \label{psi2}
&&\Psi(x_1,x_2)\phantom{,x_3,x_4}= e^{i \veck_{12}\vecx} 
  - e^{i(\theta_{21}+\veck_{21}\vecx)}, \\
\label{psi3}
&&\Psi(x_1,x_2,x_3)\phantom{,x_4}= e^{i\veck_{123}\vecx} -
  e^{i(\theta_{32}+\veck_{132}\vecx)} -
  e^{i(\theta_{21}+\veck_{213}\vecx)}\nonumber \\
  && + e^{i(\theta_{31}+\theta_{21}+\veck_{231}\vecx)} 
+ e^{i(\theta_{31}+\theta_{32}+\veck_{312}\vecx)}-
  e^{i(\theta_{32}+\theta_{31}+\theta_{21}+\veck_{321}\vecx)}, \\
  \label{psi4}
&&\Psi(x_1,x_2,x_3,x_4) = 
  e^{i\veck_{1234}\vecx}
  -e^{i(\theta_{21}+\veck_{2134}\vecx)} 
  -e^{i(\theta_{32}+\veck_{1324}\vecx)} 
  -e^{i(\theta_{43}+\veck_{1243}\vecx)}
  \nonumber \\
 &&
 +e^{i(\theta_{43}+\theta_{21}+\veck_{2143}\vecx)}
 +e^{i(\theta_{31}+\theta_{21}+\veck_{2314}\vecx)}
 +e^{i(\theta_{31}+\theta_{32}+\veck_{3124}\vecx)}
 +e^{i(\theta_{42}+\theta_{32}+\veck_{1342}\vecx)} 
 \nonumber\\
 &&
 -e^{i(\theta_{32}+\theta_{31}+\theta_{21}+\veck_{3214}\vecx)}
 -e^{i(\theta_{41}+\theta_{31}+\theta_{21}+\veck_{2341}\vecx)} 
 -e^{i(\theta_{42}+\theta_{31}+\theta_{32}+\veck_{3142}\vecx)} 
 \nonumber\\
 &&+e^{i(\theta_{41}+\theta_{32}+\theta_{31}+\theta_{21}+\veck_{3241}\vecx)}
 +e^{i(\theta_{42}+\theta_{43}+\veck_{1423}\vecx)}
 - e^{i(\theta_{43}+\theta_{42}+\theta_{32}+\veck_{1432}\vecx)}
\nonumber\\
&&- e^{i(\theta_{41}+\theta_{43}+\theta_{21}+\veck_{2413}\vecx)}
+ e^{i(\theta_{43}+\theta_{41}+\theta_{31}+\theta_{21}+\veck_{2431}\vecx)} 
+e^{i(\theta_{41}+\theta_{42}+\theta_{31}+\theta_{32}+\veck_{3412}\vecx)}
 \nonumber \\
 && -e^{i(\theta_{42}+\theta_{41}+\theta_{32}+\theta_{31}+\theta_{21}+\veck_{3421}\vecx)}
 - e^{i(\theta_{41}+\theta_{42}+\theta_{43}+\veck_{4123}\vecx)} +
  e^{i(\theta_{41}+\theta_{43}+\theta_{42}+\theta_{32}+\veck_{4132}\vecx)}
\nonumber\\
  &&+  e^{i(\theta_{42}+\theta_{41}+\theta_{43}+\theta_{21}+\veck_{4213}\vecx)}
 -e^{i(\theta_{42}+\theta_{43}+\theta_{41}+\theta_{31}+\theta_{21}+\veck_{4231}\vecx)}
\nonumber\\
&&-e^{i(\theta_{43}+\theta_{41}+\theta_{42}+\theta_{31}+\theta_{32}+\veck_{4312}\vecx)}
+e^{i(\theta_{43}+\theta_{42}+\theta_{41}+\theta_{32}+\theta_{31}+\theta_{21}+\veck_{4321}\vecx)},
\ES
in the region $0 \leq x_1 \leq \ldots \leq x_N \leq \ell$.

In the thermodynamic limit it is possible to switch from a discrete
distribution of $k$ values to a continuous distribution and the ground state
energy can then be written as \cite{Bethe31,Lieb63I}
\begin{eqnarray}
  \label{e_gamma}
  E &=& \tfrac{1}{2} N n^2\, e(\gamma)
\end{eqnarray}
where $e(\gamma)$ is given in terms of the solutions of the Lieb-Liniger
equations
\begin{eqnarray}
  e(\gamma)&=& \frac{\gamma^3}{\lambda^3(\gamma)}\int_{-1}^1 {\rm d}\xi\,
  h(\xi,\gamma)\xi^2 \\
  \label{LLeqs}    
  h(\xi,\gamma)&=&\frac{1}{2\pi}+\frac{1}{\pi}\int_{-1}^1 {\rm d}y\,
  \frac{\lambda(\gamma) h(y,\gamma)}{\lambda^2(\gamma) + (y-\xi)^2},\quad
  \lambda(\gamma) = \gamma \int_{-1}^1 {\rm d}\xi\,
  h(\xi,\gamma).
\end{eqnarray}
Using the Hellmann--Feynman theorem \cite{Feynman39}, we can directly obtain
the second order correlation function along the diagonal from the ground state
energy
\begin{eqnarray}
  g^{(2)}(x,x)& \equiv& g^{(2)}(0,0)\; = \; 
  \frac{\partial}{\partial \gamma}e_{B}(N,\gamma),
\end{eqnarray}
which can be measured experimentally
\cite{Ertmer03,Weiss05,shimitsu96,Kasevich,CohenTannoudji97,westbrook07}.  In
the thermodynamic limit, this simplifies to $g^{(2)}(x,x)\equiv
g^{(2)}(0,0)=e'(\gamma)$.
\begin{figure}[ht]
  \centering
  \includegraphics[width=0.8\columnwidth]{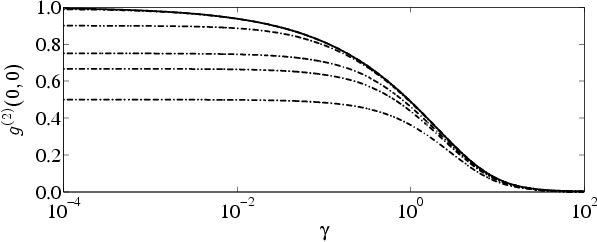}
\caption{Second order correlation 
  function $g^{(2)}(0,0)$ vs. $\gamma$ for a homogeneous setup. Results from
  the thermodynamic limit (solid line) are compared to -- from bottom to top
  -- results for $N=2,3,4,10,100$ bosons (dash-dotted lines).}
\label{Comp_LL_e}
\end{figure}

\section{Correlation functions of the homogeneous system}\label{hom}
To get a feeling for the second order correlation function and its dependence
on the particle number $N$ as well as the correlation strength $\gamma$, we
first consider the homogeneous system, where this is straight forward.  The
results are depicted in figure~\ref{Comp_LL_e} where we compare $g^{(2)}(0,0)$
in the thermodynamic limit to results that are obtained by using finite values
of the particle number $N$. As can easily be seen, the thermodynamic limit is
approached very quickly and for $N=100$ bosons the exact results obtained with
the Bethe ansatz are almost indistinguishable from the thermodynamic limit.

The decrease of the second order correlation function for large values of
$\gamma$ is due to an interaction induced antibunching also known as
fermionization of bosons. In the limit $\gamma \to \infty$, this behavior was
first predicted by Girardeau \cite{Gir60} and is due to a one-to-one
correspondence between impenetrable bosons and spinless fermions in
one-dimensional systems. This can be understood by looking at the limiting
values of the wave vector $\veck$ and the phases $\theta_{rs}$ as a function of the
interaction parameter $\gamma$. At large values of $\gamma$, the wave functions
of (\ref{psi2}), (\ref{psi3}) and (\ref{psi4}) approach the form of
Slater determinants. Thus, the wave-function of $N$ bosonic particles must
vanish when two particles are getting close as if they were fermions.
\begin{figure}[ht]
  \centering
  \includegraphics[width=0.8\columnwidth]{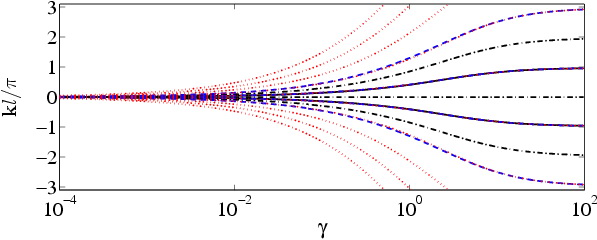}
  \caption{Wave vector 
    $\veck\ell/\pi$ vs. $\gamma$ for $N=2$ (solid lines), $N=3$ (dash-dotted
    lines), $N=4$ (dashed lines) and $N=10$ (dotted lines). The functional
    form of the wave vectors is very similar for different particle numbers,
    there is only a different number of wave vector components for each
    particle number.}
  \label{wave_vectors}
\end{figure}

In figure~\ref{wave_vectors} we plot the results for the wave vectors for an
increasing particle number $N=2,3,4,10$. For odd particle numbers the
components of the wave vectors approach even multiples of $2\pi$, which is
exactly the result that one would obtain for non-interacting, spinless
fermions subject to periodic boundary conditions. However for even particle
numbers, the components of the wave vectors are also separated by a constant
spacing of $2 \pi$ if we go to strong interactions, but they approach odd
multiples of $\pi$.

This does not correspond to the behavior of
non-interacting, spinless fermions subject to periodic boundary conditions.
This breakdown of the mapping between impenetrable bosons and spinless
fermions was already pointed out by Girardeau and originates from the periodic
boundary conditions. Considering hard wall boundary conditions,
then there are no restrictions on the particle number $N$.  Furthermore it can
be noticed that the dependence of the wave vectors on the correlation
parameter $\gamma$ is very similar for the different particle numbers. Hence
the two components of the wave vectors with smallest magnitude for $N=4$ and
$N=10$ are indistinguishable from the solid line which depicts the components
of the wave vector for $N=2$.

Apart from the behavior of the wave vectors themselves, a closer look on the
wave function reveals that the factors $a_\mathcal{P}$ appearing in
\eqref{ansatz} approach either $+1$ or $-1$ for $\gamma \to \infty$. Hence the
wave function approaches the limit of a Slater determinant as introduced by
Girardeau.  A simple way of understanding these behaviors is to consider the
defining equations for the wave vectors \eqref{quotient1} and
\eqref{quotient2} in the limit $\gamma \to \infty$. Using a linear
approximation for the $\arctan$, one obtains \BS
\label{gamma_inf}
(k_{j+1}-k_j)\ell &=& 2 \pi - \frac{4\pi}{\gamma} + O(1/\gamma^2) \quad , \quad
\theta_{lj} =\frac{4\pi(j-l)}{N \gamma} + O(1/\gamma^2) \ES which
exactly yield the previously explained behavior for $\gamma \to \infty$.

In the remainder of this section we want to analyze the spatial dependence of
the second order correlation function for the homogeneous problem. Due to the
translational symmetry of the system, the correlation functions only depend on
the relative distance $|x-y|$. Hence, it suffices to just study the
anti-diagonal of the correlation functions for $y=-x$. In particular, we will
examine the particle numbers $N=2,3,4$, because they exhibit finite number
effects which vanish in the thermodynamic limit \cite{Schmidt07}.
\begin{figure}
  \centering
  \includegraphics[width=0.32\columnwidth]{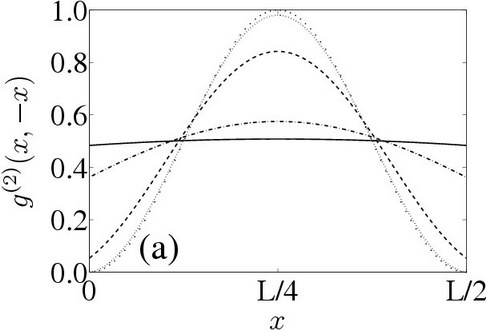}
  \includegraphics[width=0.32\columnwidth]{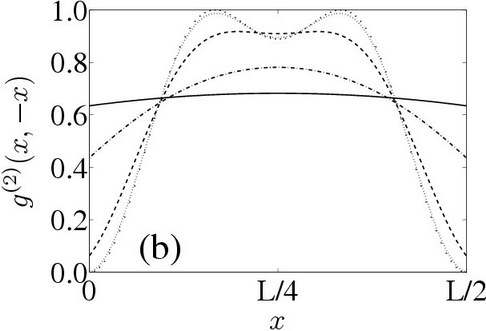}
  \includegraphics[width=0.32\columnwidth]{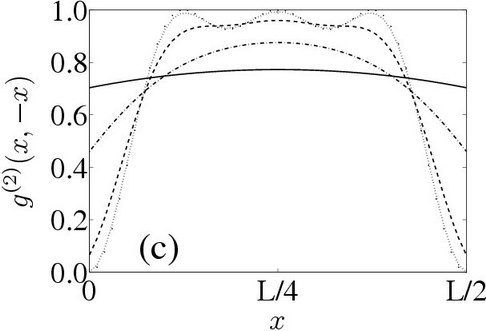}
\caption{Anti-diagonal of the second order 
  correlation function $g^{(2)}(x,-x)$ vs. position $x$ for the homogeneous
  system.  Results for $N=2$ are depicted in subplot a) for $N=3$ in subplot
  b) and for $N=4$ in subplot c).  The lines correspond to values of the
  correlation parameter of $\gamma=0.1$ (solid line), $\gamma=1$
  (dashed-dotted line), $\gamma=10$ (dashed line), $\gamma=100$ (thin dotted
  line) and $\gamma=\infty$ (thick isolated dots).}
\label{g2_Bethe_N3}
\end{figure}
In figure~\ref{g2_Bethe_N3} we depict the anti-diagonal of the second order
correlation function $g^{(2)}(x,-x)$ for $N=2,3,\text{ and }4$ particles,
respectively. For an increasing value of the correlation parameter $\gamma$,
we again notice the strong antibunching at the origin, which reduces
$\lim_{\gamma\rightarrow 0}g^{(2)}(0,0,\gamma)=1-1/N$ for ideal bosons and
$\lim_{\gamma\rightarrow\infty} g^{(2)}(x,-x,\gamma)=0$ .

Furthermore the off-diagonal develops an oscillatory behavior which becomes
most pronounced for large values of the correlation parameter.  The
oscillatory behavior strongly depends on the particle number.  For $N$
strongly interacting particles in the system we expect $N-1$ maxima of
$g^{(2)}(x,-x)$ in the interval $0\leq x \leq L/2$.  This can be understood by
recalling the behavior of the wave vectors. In the limit $\gamma \to \infty$
their components are all equally separated by $2\pi$ and therefore exactly
$N-1$ different combinations of the components exist in the calculation of
$g^{(2)}(x,-x)$ when we insert \eqref{ansatz} in \eqref{g2}. These
combinations range from $2\pi$ to $(N-1)2\pi$ and lead to trigonometric
functions with $N-1$ different frequencies, thereby producing the behavior
that was described above. Finally, it is also interesting to note that the
anti-diagonal of the second order correlation function has a kink at $x=0$
which is due to the delta interaction of the particles. However, in the limit
$\gamma \to \infty$ the kink vanishes and $g^{(2)}(x,-x)$ approaches a smooth
behavior at the origin.

In general, it is possible to write down the analytic form of any correlation
function because we have knowledge of the complete many-particle wave
function. Using the particular form of the wave function for $N=2,3,4$, as
given in ~\eqref{psi2}, \eqref{psi3} and \eqref{psi4}, we have calculated
the second order correlation function according to \eqref{g2} by properly
using the total symmetry of the respective wave functions. However, the
lengthy analytical results are not enlightening and we refrain from writing
them down in detail.  Nevertheless, studying them in the limit $\gamma \to
\infty$, using \eqref{gamma_inf}, we find the following simple result\BS
g^{(2)}(x,-x)&=& \frac{N-1}{N} \left( 1 - \sum\limits_{j=1}^{N-1}
  \frac{2(N-j)}{N(N-1)}\cos\Big(j\cdot 2\pi \frac{2x}{\ell}\Big)\right) +O(
1/\gamma)\; ,  \ES where we have also extrapolated the results for $N=2,3,4$
to general particle numbers $N$. This is consistent with the known limit for
non-interacting fermions, as can be seen by evaluating the geometric series
\BS g^{(2)}(x,-x)&=& 1 - \frac{\sin^2(2\pi x N/\ell)}{N^2 \sin^2(2\pi x
  /\ell)} \; . \ES This result was already obtained by M.  Girardeau
\cite{Gir60}.  However he first considered the limit $\gamma \to \infty$ and
used the fermionic wave function for the evaluation of the second order
correlation function.


\section{Correlation functions of the inhomogeneous system}\label{inhom}
\begin{figure}[ht]
  \centering 
  \includegraphics[angle=-90,width=0.32\columnwidth]{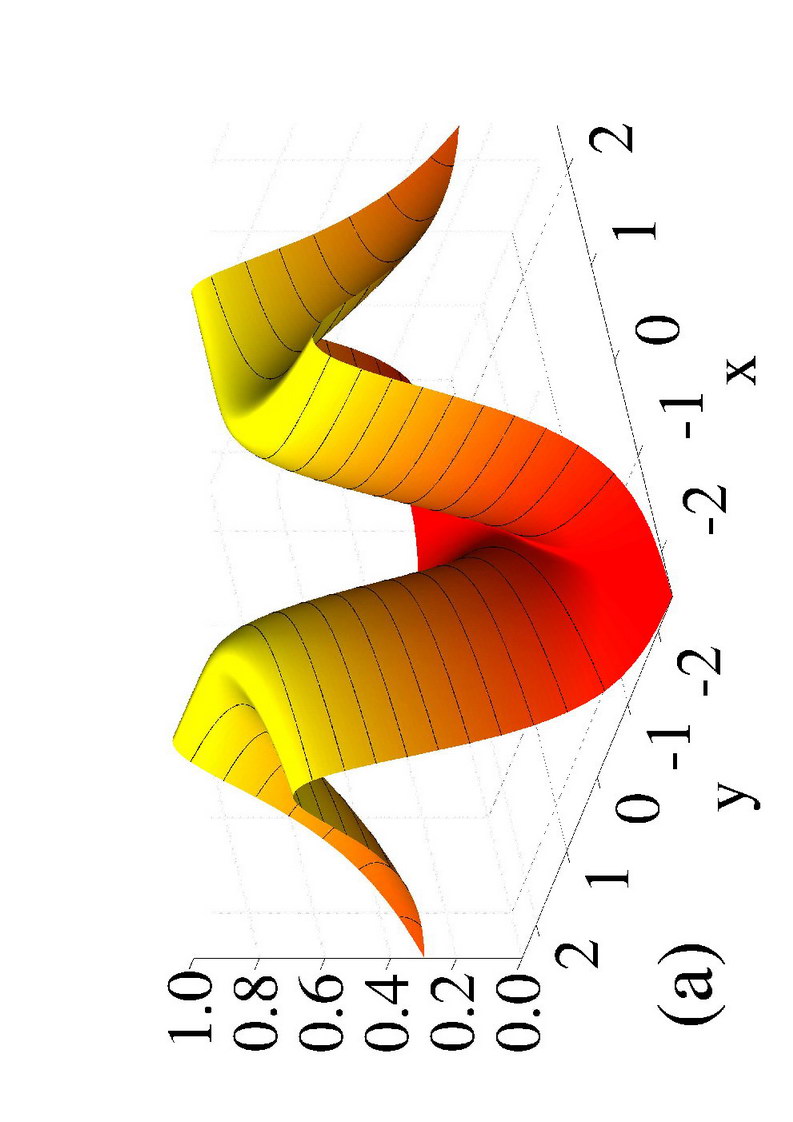}
  \includegraphics[angle=-90,width=0.32\columnwidth]{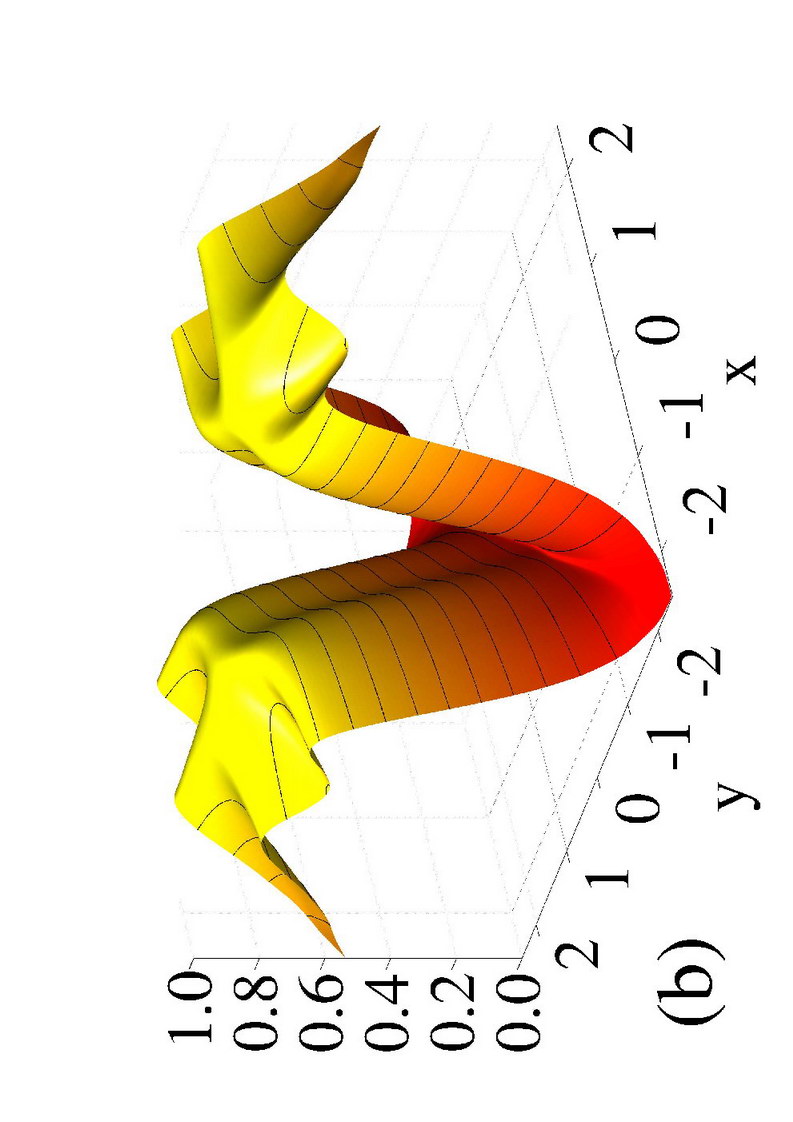}
  \includegraphics[angle=-90,width=0.32\columnwidth]{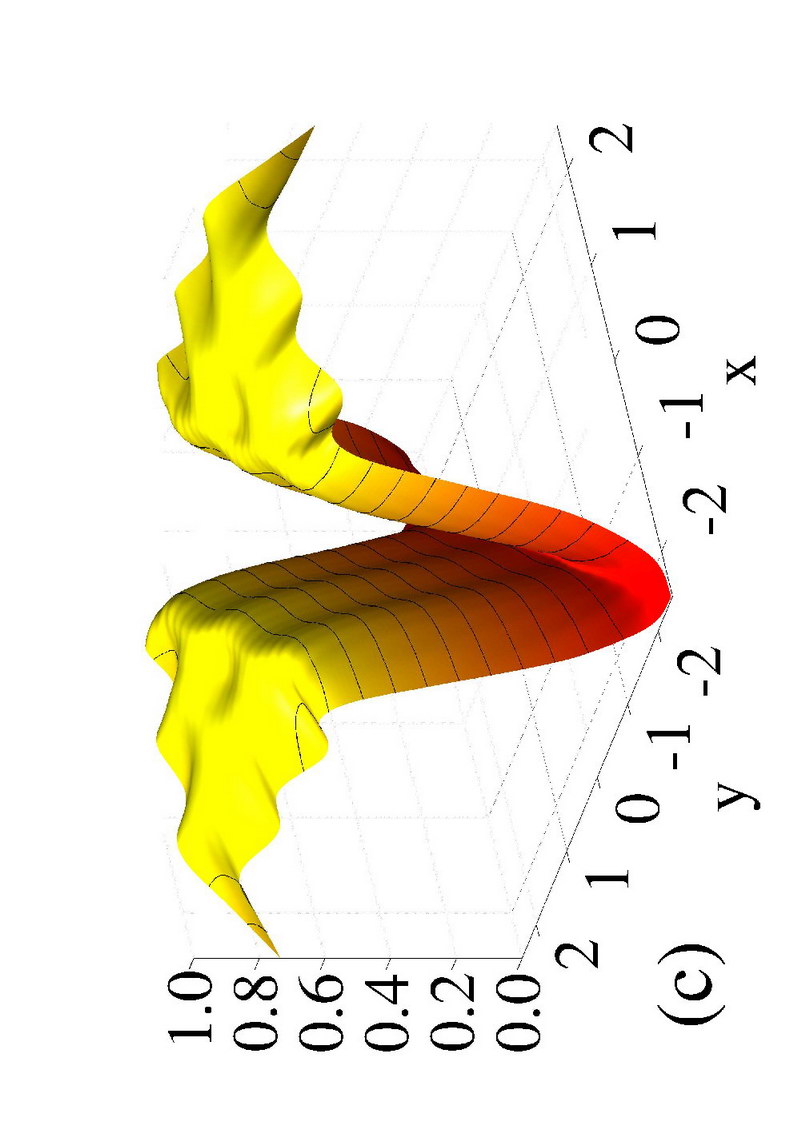}
  \caption{Second order correlation function 
    $g^{(2)}(x,y)$ for harmonically trapped particles versus the two-particle
    coordinate $x,y$ for $N=2$ in subplot a), for $N=3$ in subplot b) and for
    $N=4$ in subplot c). The interaction strength is $g=10$.}
  \label{pic_g2_N3_N4}
\end{figure}

With this understanding of the second order correlation function and its
dependence on the particle number $N$ and correlation strength $\gamma$, we
are now in the position to interpret the trapped results. For $N=2$ particles,
exact results for the ground state wave function are known
\cite{Englert98,cirone} and for $N=3,4$ particles we used the MCTDH method.
The corresponding results for the spatial correlations and a fixed interaction
strength of $g=10$ are shown in figure~\ref{pic_g2_N3_N4}.

\begin{figure}[ht]
  \centering
  \includegraphics[width=0.8\columnwidth]{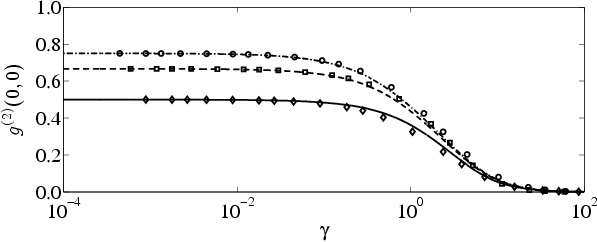}
\caption{Second order correlation function $g^{(2)}(0,0)$ in 
  the center of the trap as a function of the correlation parameter $\gamma$.
  The data corresponds to interaction strengths ranging from $g=0.001$ to
  $g=50$. The results for $N=2,3,4$ trapped particles (diamonds, squares,
  circles) are compared to a homogeneous system (solid, dashed, dashed-dotted
  line).}
\label{Comp_N2_N3_N4}
\end{figure}

To establish a common ground for the comparison of the homogeneous and the
trapped system, we first of all restrict ourselves to the values of the second
order correlation function in the center of the trap $g^{(2)}(x=0,0)$.  Tuning
the interaction from $g=0.001 \rightarrow 50$, we cover the whole range of the
correlation parameter $\gamma$ from the weakly interacting GP regime to the TG
regime of strong interactions.  In figure \ref{Comp_N2_N3_N4} we compare the
results in the trapped system for $N=2,3,4$ particles (diamonds, squares and
circles) to the homogeneous results (solid, dashed and dashed-dotted line) for
the same parameters of $g$ and $N$. The length of the homogeneous system
$\ell$ is chosen such that the constant number density $n=N/\ell$ equals the
number density of the trapped system in the center of the trap. In general we
can see a good agreement of the results with slight deviations at the
crossover from weakly to strongly interacting bosons around $\gamma=1$.

Thus we conclude that the correlation parameter $\gamma$ remains a valid
parameter for the description of the inhomogeneous system as well. It is
therefore possible to use a position dependent correlation parameter
$\gamma(x)=g/n(x)$, which is the central hypothesis of the local density
approximation (LDA).

\begin{figure}[ht]
  \centering 
  \includegraphics[width=0.32\columnwidth]{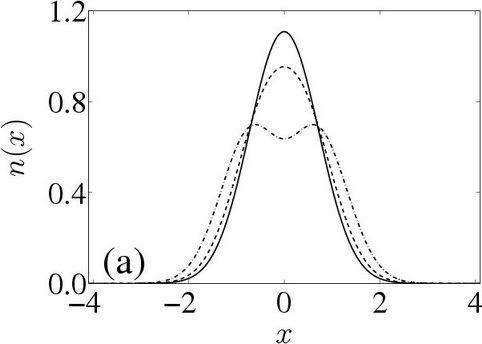}
  \includegraphics[width=0.32\columnwidth]{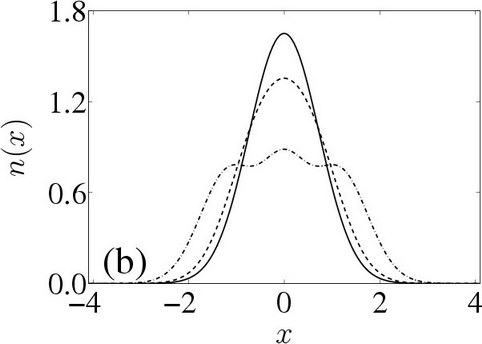}
  \includegraphics[width=0.32\columnwidth]{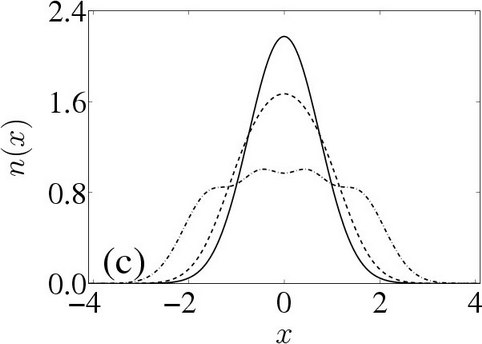}
  \caption{Number density $n(x)$ for $N=2$ 
    in subplot a), for $N=3$ in subplot b) and for $N=4$ in subplot c). The
    data corresponds to interaction strengths of $g=0.1$, $g=1$ and $g=10$
    (solid, dashed, dashed-dotted line).}
\label{den_N2_N3_N4}
\end{figure}

In further considerations, we will apply this position dependent correlation
parameter. Thus it is necessary to be acquainted with its spatial behavior.
Hence, we plot the number density $n(x)$ for $N=2,3,4$ particles for various
strengths of the interaction in figure~\ref{den_N2_N3_N4}. With an increasing
interaction strength the density of the ground state changes from a Gaussian
shape at weak interactions to a broadened distribution with Friedel
oscillations \cite{Friedel58} for strong interactions.

From the previous discussion we learn that the second order correlation
function in the center of the trap, $g^{(2)}(0,0)$, can well be understood by
looking at the corresponding correlation function in the homogeneous system
for the same parameters of $g$, $N$ and $\gamma$. In the next step we want to
investigate the behavior of the diagonal of the second order correlation
function, $g^{(2)}(x,x)$. For this purpose we take the exact values of the
diagonal of the second order correlation function for the trapped system
(obtained with an MCTDH calculation) and plot them as a function of the
position dependent correlation parameter $\gamma(x)$. Comparing these values
with the diagonal behavior of the second order correlation function in the
homogeneous case, presented in figures~\ref{Comp_LL_e} and \ref{Comp_N2_N3_N4},
basically corresponds to a local density approximation. As the density
decreases by moving out of the center of the trap, increasing values of
$\gamma$ correspond to increasing values of $x$.
\begin{figure}[ht]
  \centering
  \includegraphics[width=0.49\columnwidth]{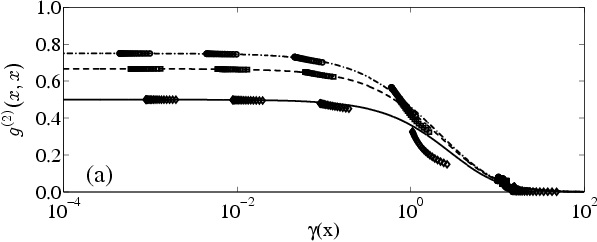}
  \includegraphics[width=0.49\columnwidth]{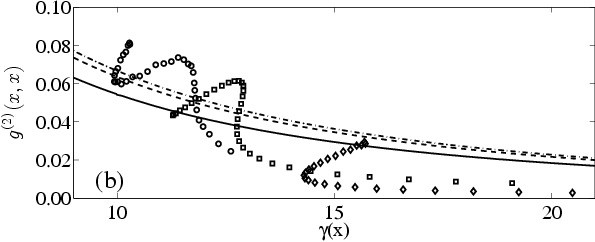}
\caption{Diagonal of the second order correlation function $g^{(2)}(x,x)$ as
  a function of the correlation parameter $\gamma(x)$. The data corresponds to
  interaction strengths of $g=0.001$, $g=0.01$, $g=0.1$, $g=1$ and $g=10$. The
  results for $N=2,3,4$ for the trapped system (diamonds, squares, circles)
  are compared to the results for the homogeneous case (solid, dashed,
  dashed-dotted line) over the whole range of the correlation parameter in
  subplot a). A magnification of the behavior of the second order correlation
  function around $\gamma =15$ is shown in subplot b).}
\label{Comp_N2_N3_N4_diag}
\end{figure}

In figure~\ref{Comp_N2_N3_N4_diag} we plot the diagonal of the second order
correlation function $g^{(2)}(x,x)$ for $N=2,3,4$ at interaction strengths of
$g=0.001$, $g=0.01$, $g=0.1$, $g=1$ and $g=10$. Generally speaking, the
results for $N=2,3,4$ for the trapped system (diamonds, squares, circles)
agree rather well with the results of the Bethe ansatz for the homogeneous
system in the weakly interacting regime. However the trapped system deviates
significantly from the homogeneous results for large correlation parameters.
Starting with values of the correlation parameter around $\gamma\approx 1$, we
can see oscillations of the trapped results around the homogeneous curve. This
means that the local density approximation gets less suited to describe the
physical content of the trapped system. 

The breakdown of the local density
approximation can well be understood if one considers the behavior of the
density in the trap. For large interaction strengths the density develops the
previously mentioned Friedel-type oscillations which can not be described
within a local density approximation. This translates to the second order
correlation function where similar oscillations occur.

Whereas for $N=3$ the Friedel type oscillations lead to a peak of the density
in the center of the trap, the opposite is true for $N=2,4$, where we have a
dip in the center. In terms of the second order correlation function this
leads to oscillations around the homogeneous result that either start below
the homogeneous curve ($N=3$) or above it ($N=2,4$), as one moves out of the
center of the trap. This can be seen in subplot b) of
figure~\ref{Comp_N2_N3_N4_diag} which magnifies the behavior of the second order
correlation function around $\gamma =15$.
\begin{figure}[ht]
  \centering
  \includegraphics[angle=-90,width=0.32\columnwidth]{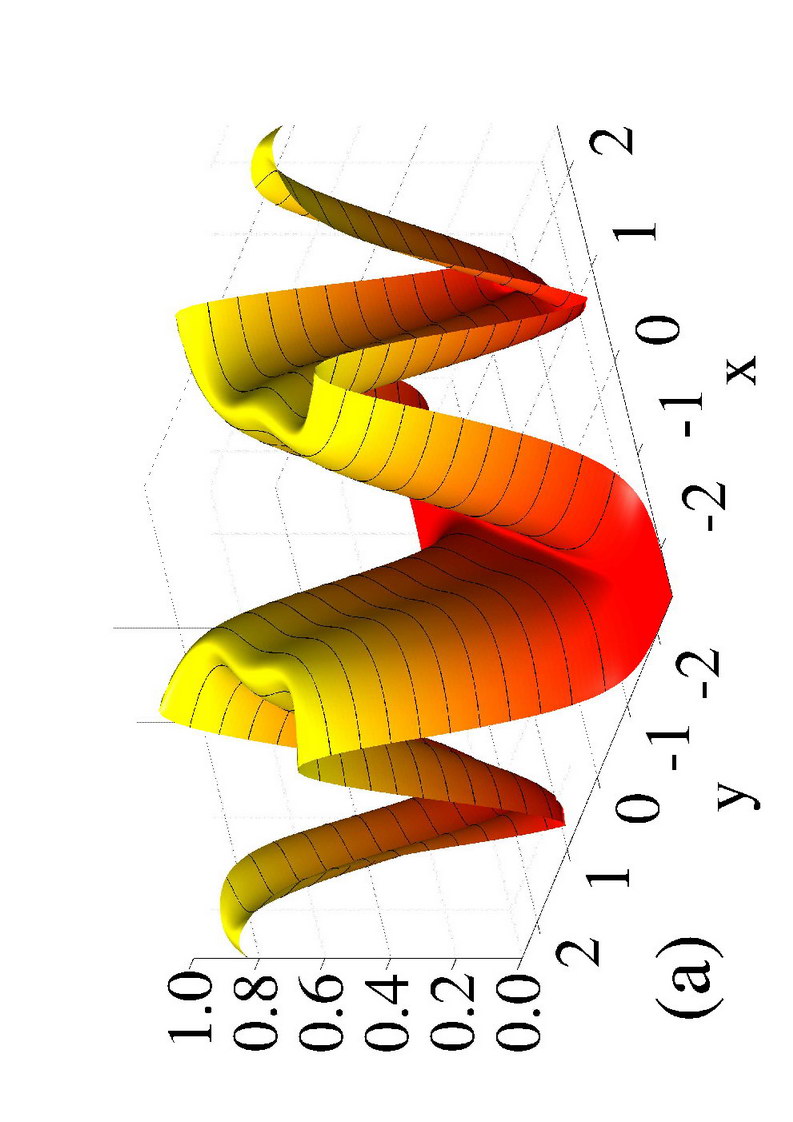}
  \includegraphics[angle=-90,width=0.32\columnwidth]{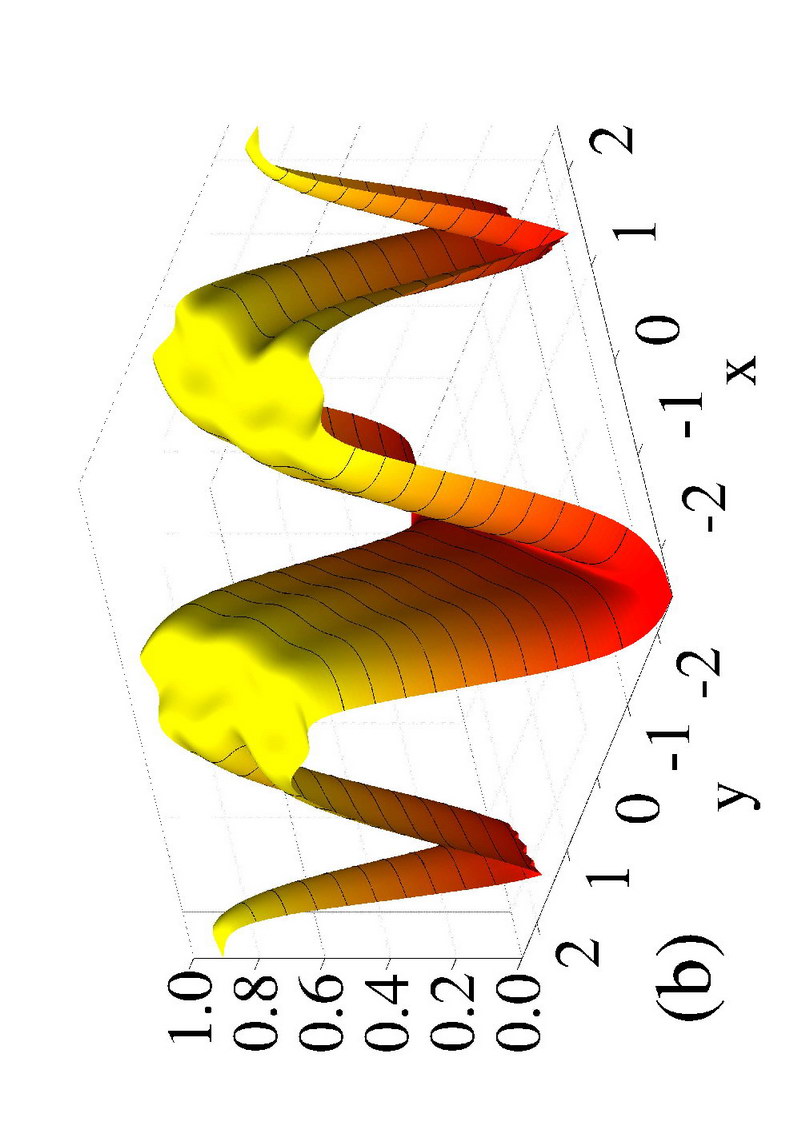}
  \includegraphics[angle=-90,width=0.32\columnwidth]{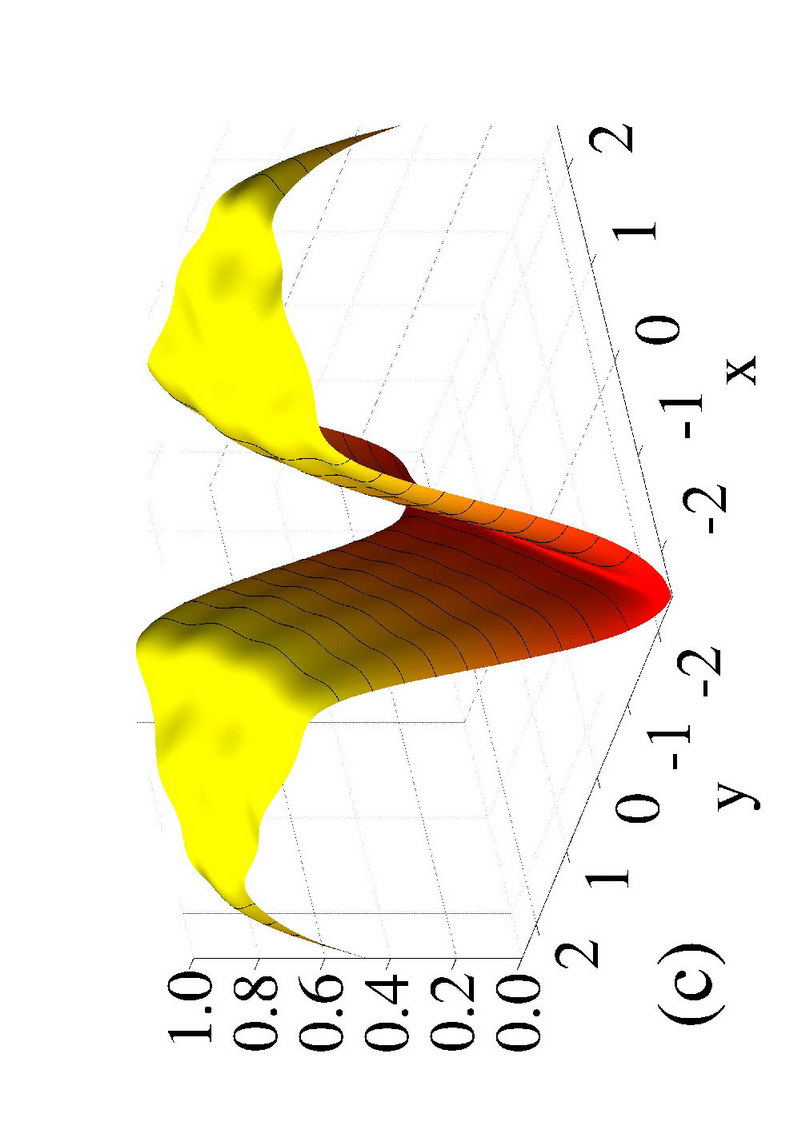}
\caption{Second order correlation function $g^{(2)}(x,y)$ 
  in a local density approximation combined with the Bethe ansatz for $N=2$ in
  subplot a), for $N=3$ in subplot b) and for $N=4$ in subplot c).  The
  interaction strength is in each case $g=10$.}
\label{pic_g2_N3_N4_Bethe}
\end{figure}

Apart from an oscillatory behavior for large interaction strengths, we have
seen that the diagonal of the trapped system can be understood if we combine
the Bethe ansatz with a local density approximation, in the sense described
above. Finally, we want to investigate to which extent the full behavior of
$g^{(2)}(x,y)$ can be analyzed by the same means.  

In this last step of the
comparison we calculate the value of $g^{(2)}(x,y)$ in the local density
approximation in the following way. In analogy to the previous discussion, we first of all 
recall that the comparison of the homogeneous and the inhomogeneous system is 
made for the same values of the particle number $N$ and the coupling constant $g$. 
In the original homogeneous system we have the translational symmetry and the 
properties of the correlation functions only depend on the relative distance $|x-y|$.
In an inhomogeneous system the relative distance is not the only relevant property that 
determines the behavior of the second order correlation function $g^{(2)}(x,y)$.
Instead we have to incorporate the spatial dependence of the density to arrive at a
local density approximation. For the diagonal part of the correlation function we have already seen that this combination of the Bethe ansatz and the local density approximation leads to a good agreement with the results for the inhomogeneous system. We can extend this procedure to non-diagonal coordinate pairs $(x,y)$ by choosing the
density at the center of mass coordinate $(x+y)/2$ for the local density approximation. As previously, this density is given by the exact MCTDH results and in the next step we again adjust the dimensionless length $l$ for the homogeneous system, such that we obtain the same density for the fixed particle number $N$. Thereafter we solve the Bethe ansatz for these parameters and extract the anti-diagonal value of the second order correlation function for a relative distance of $|x-y|$. This result is eventually used to represent $g^{(2)}(x,y)$ for the combination of the Bethe ansatz and the local density approximation. This procedure is repeated for every coordinate pair $(x,y)$ which is used to plot the second order correlation function.

In this fashion we
obtain the counterparts to the exact trapped results in figure
\ref{pic_g2_N3_N4} and depict them in figure~\ref{pic_g2_N3_N4_Bethe}. The most
striking difference that can be noticed by comparing the homogeneous to the
trapped results is the large dip in the off-diagonal for $N=2,3$ in the
homogeneous case with local density approximation. This is merely due to the
fact that the periodic boundary conditions in the Bethe ansatz prevent one
from going to large distances. For a larger particle number and consequently a
larger number density this difference begins to be negligible in the region of
interest, as can be seen in the plot for $N=4$. Apart from this artifact the
general features are similar and we conclude that we can also understand the
overall behavior of the second order correlation function in terms of a local
density approximation and the Bethe ansatz.

\begin{figure}[ht]
  \centering
  \includegraphics[width=0.32\columnwidth]{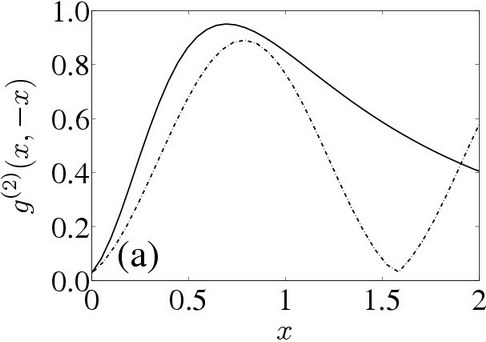}
  \includegraphics[width=0.32\columnwidth]{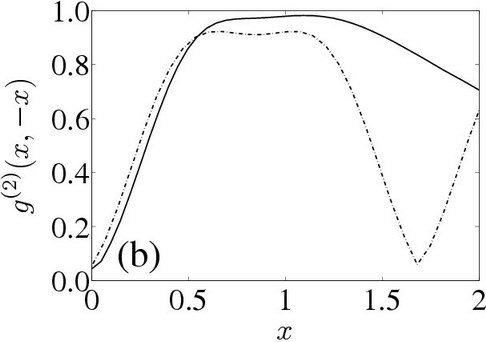}
  \includegraphics[width=0.32\columnwidth]{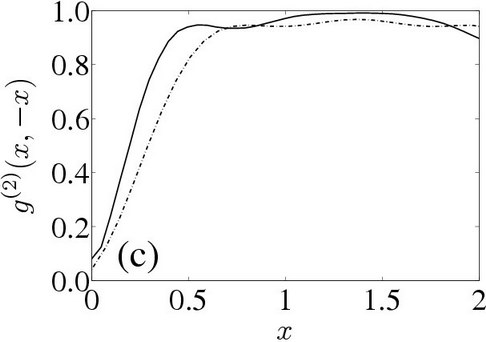}
\caption{Anti-diagonal of the second order 
  correlation function $g^{(2)}(x,-x)$ across the center of the trap for $N=2$ in
  subplot a), for $N=3$ in subplot b) and for $N=4$ in subplot c). The
  interaction strength is $g=10$ and the MCTDH results (solid line) are
  compared to the combination of a local density approximation and the Bethe
  ansatz (dashed-dotted line).}
\label{Comp_g2_anti_N3}
\end{figure}

Having a closer look at the anti-diagonal in the center of the trap in
figure~\ref{Comp_g2_anti_N3} we get a clearer illustration of these conclusions.
While the periodicity prevents one from matching the exact physical behavior
for any $x$, at least the initial slope and the approximate shape for $x<1$
are modeled well.


\section{Conclusion}\label{con}
We have examined the ground state correlations for repulsive, quasi
one-dimensional bosons in a harmonic trap. In particular, we have focused on
the few particle limit $N=2,3,4,\ldots$, where exact numerical solutions of
the many particle Schr\"odinger equation are available with the
Multi-Configuration Hartree method.  These numerical results for the
inhomogeneous system are modeled with the analytical solution of the
homogeneous problem using the Bethe ansatz and the local density
approximation.  Tuning the interaction strength from the weakly correlated
Gross-Pitaevskii- to the strongly correlated Tonks-Girardeau regime reveals
finite number effects in the second order correlation function beyond the
local density approximation.

\section*{Acknowledgments}
ME, RW and WPS acknowledge the financial support of this work by the
German Science Foundation via project B5 of the SFB/TRR 21. 
Financial support from the Landesstiftung Baden-
W\"urttemberg through the project "Mesoscopics and atom
optics of small ensembles of ultracold atoms" is gratefully
acknowledged by PS and SZ.

\bibliography{References}

\end{document}